\documentclass[aps,eqsecnum,groupedaddress,fleqn]{revtex4}
\usepackage{graphicx}
\usepackage{amssymb}
\newcommand{\be}{\begin{equation}}
\newcommand{\ee}{\end{equation}}
\newcommand{\bea}{\begin{eqnarray}}
\newcommand{\eea}{\end{eqnarray}}
\newcommand{\bvec}[1]{\mbox{\boldmath $#1$}}
\newcommand{\eqref}[1]{(\ref{#1})}
\begin{document}


\preprint{KANAZAWA-02-18}

\title{Generalized flux-tube solution in Abelian-projected SU($N$) gauge theory}

\author{Yoshiaki Koma}
\email[]{koma@hep.s.kanazawa-u.ac.jp}
\affiliation{Institute for Theoretical Physics, Kanazawa University,\\
Kakuma-machi, Kanazawa, Ishikawa 920-1192, Japan}

\date{\today}

\begin{abstract}
The [U(1)]$^{N-1}$ dual Ginzburg-Landau (DGL) theory 
as a low-energy effective theory of Abelian-projected SU($N$) gauge theory
is formulated in a Weyl symmetric way. 
The string tensions of flux-tube solutions of the DGL theory 
associated with color-electric charges in various representations of SU($N$) 
are calculated analytically at the border between type-I and type-II 
of the dual superconducting vacuum (Bogomol'nyi limit).
The resulting string tensions satisfy the flux counting rule, which
reflects the non-Abelian nature of gauge theory.
\end{abstract}

\maketitle

\section{Introduction}

\par
The dual superconducting picture of the QCD vacuum
provides an illuminating scenario of the quark confinement 
mechanism in QCD~\cite{tHooft:1975pu,Mandelstam:1974pi}.
For example, in this vacuum the color-electric flux associated with 
a quark-antiquark system is squeezed into an 
almost-one-dimensional object due to the dual Meissner effect,
which forms the so-called color-electric flux tube.
If the flux tube is long enough it acquires a constant energy 
per unit length which we call the string tension. 
A linearly  rising confinement potential has naturally appeared.
This picture is analytically described by the
dual Ginzburg-Landau (DGL)
theory~\cite{Suzuki:1988yq,Maedan:1988yi,Suganuma:1995ps,Sasaki:1995sa}.

\par
The DGL theory is constructed from QCD,
following 't Hooft, by performing an Abelian 
projection~\cite{tHooft:1981ht}.
Two hypotheses enter this construction, Abelian 
dominance and monopole condensation~\cite{Ezawa:1982bf,Ezawa:1982ey}.
These hypotheses are numerically supported by many studies of 
SU(2) and SU(3) lattice gauge theories in 
the maximally Abelian gauge (MAG)~\cite{Kronfeld:1987vd,%
Suzuki:1990gp,Brandstater:1991sn,Shiba:1995db,Bali:1996dm}.
The importance of monopoles, of Abelian degrees of freedom
in general, and the resulting dual superconducting properties 
have been confirmed even in full QCD 
in maximally Abelian gauge (MAG)~\cite{Bornyakov:2001nd}.
In view of these observations, the dual superconductor picture
seems to be realized as an universal feature of the vacuum
in arbitrary SU($N$) gauge theory.
As for $N \geq 3$, this could be expected since SU($N\geq 3$) 
contains SU(2) as a subgroup.

\par
In the present paper, we elaborate on the dual superconducting scenario
applied to the SU($N$) (with $N>3$) gauge theory by formulating explicitly
the [U(1)]$^{N-1}$ DGL theory under the usual assumptions of 
Abelian dominance and monopole condensation.
Based on the Weyl symmetric 
formulation~\cite{Koma:2000wn,Koma:2001ut,Koma:2000hw}, 
we are able to simplify the DGL framework and calculate systematically 
the string tensions of flux tubes associated with 
color-electric charges in various 
representations of SU($N$).
Finally, we extrapolate the behavior of the resulting string tensions
to the large $N$ limit.

\par
Our main interest here is to learn
how the original non-Abelian SU($N$) gauge symmetry
is reflected in the systematics of the flux-tube solutions
that emerges  in the [U(1)]$^{N-1}$  DGL theory.
In order to be able to work out these consequences in an analytically
accessible case first, it is recommendable to start these considerations
at the point separating between the type-I and the type-II dual 
superconducting vacuum, at the so-called Bogomol'nyi 
limit~\cite{Bogomolny:1976de,deVega:1976mi,%
Chernodub:1999xi,Koma:2000wn},
where the {\it flux counting rule} will become manifest.
The flux counting rule indicates that the string tension of the flux tube  
is proportional to the topological winding number of the flux, 
which depends on the color-electric charges attached to its ends.
It is instructive to compare the resulting rule with the 
eigenvalues of quadratic Casimir operator in various 
representations, to which the different 
color charges of the non-Abelian gauge theory can belong.
We shall make some important observations at this point.

\section{The [U(1)]$^{N-1}$ Dual Ginzburg-Landau Theory}

\par
In the SU($N$) case,  the color-electric charges of quarks
and the color-magnetic charges of monopoles 
can be defined by using the weight vectors and 
the root vectors of the SU($N$) algebra as 
$\vec{Q}_{k}^{(e)}=e \vec{\omega}_{k}$ ($k=1,2,\ldots,N$),
and 
$\vec{Q}_{ij}^{(m)}=g \vec{\varepsilon}_{ij}$ ($i=1,2,\ldots,N$ and
$j=i+1,\ldots,N$), 
where 
$\vec{\varepsilon}_{ij} \equiv \vec{\omega}_{i}-\vec{\omega}_{j}$.
The Dirac quantization condition, $eg=4\pi$, is now
extended to
\bea
\vec{Q}_{ij}^{(m)}\cdot \vec{Q}_{k}^{(e)} = 2 \pi m_{ij\; k} \; ,  
\label{eq:Dirac_1}
\eea
where 
\be
m_{ij\; k} \equiv 2 \vec{\varepsilon}_{ij} 
\cdot \vec{\omega}_{k} =\delta_{ik}-\delta_{jk}
\ee
takes integer values only.

\par
In the Cartan representation, the [U(1)]$^{N-1}$ DGL theory has the 
form~\footnote{Throughout this paper, we use the notations:
Latin indices $i$ and $j$ denote color labels, which
are not to be summed over unless explicitly mentioned.
Boldface letters are used for field variables which are three-dimensional 
spatial vectors.}
\bea
{\cal L}_{\rm DGL} =
-\frac{1}{4} 
( \partial_{\mu} \vec{B}_{\nu} -
\partial_{\nu} \vec{B}_{\mu}
- e \sum_{k=1}^{N} \vec{\omega}_{k} \Sigma_{k\;\mu\nu}^{(e)} )^{2}
+
\sum_{i < j}^{N}
\left[ 
\left | ( \partial_{\mu} + i g \vec{\varepsilon}_{ij} \cdot 
\vec{B}_{\mu}) \chi_{ij} \right |^{2}
- \lambda \left( |\chi_{ij}|^{2} - v^{2} \right)^{2}   \right] \; ,
\label{eqn:Cartan-DGL}
\eea
where $\vec{B}_{\mu}=(B_{\mu}^{(3)}, 
B_{\mu}^{(8)},\ldots,B_{\mu}^{(N^{2}-1)})$
is the dual gauge field and $\chi_{ij}$ the complex-scalar
monopole field which has $N(N-1)/2$ components.
For the potential of the monopole field we adopt the most 
simple form which is able to incorporate monopole condensation.
In the presence of external color-electric charges,
a color-electric Dirac string $\Sigma_{k\;\mu\nu}^{(e)}$
($k=1,2,\ldots,N$) appears in the dual field strength tensor, 
where the color-electric charge $\vec{Q}_{k}^{(e)}$ is factorized out.
In the dual description, this Dirac string singularity is 
important to define the color-electric current $j_{k \; \mu}^{(e)}$ 
as a boundary term
$j_{k \; \mu}^{(e)}=\partial^{\nu} {}^{*\!}\Sigma_{k\;\mu\nu}^{(e)}$,
which leads to the violation of the {\it dual} Abelian Bianchi identity.
The characteristic scales are given by the masses of
the monopole field $m_{\chi}=2 \sqrt{\lambda} v$ and of
the dual gauge field $m_{B}=\sqrt{N} g v $.
The ratio between them, $\kappa \equiv m_{\chi}/m_{B}$,
is the Ginzburg-Landau parameter which classifies the type 
of the vacuum: 
$\kappa <1$ (type-I) and $\kappa >1$ (type-II).

\par
The dual field strength tensor  
${}^{*\!}\vec{F}_{\mu\nu}$ appearing in (\ref{eqn:Cartan-DGL})
can be rewritten in a way lined up with the components of the Higgs field. 
Redefining the dual gauge field as
\be
B_{ij\;\mu} \equiv g \vec{\varepsilon}_{ij} \cdot \vec{B}_{\mu} \; ,
\ee
we can rewrite the dual field strength tensor as
\bea
{}^{*\!}\vec{F}_{\mu\nu} 
&\equiv&
\partial_{\mu} \vec{B}_{\nu} -
\partial_{\nu} \vec{B}_{\mu}
- e \sum_{k=1}^{N} \vec{\omega}_{k} \Sigma_{k\;\mu\nu}^{(e)}
=
\frac{1}{Ng} \sum_{i=1}^{N} \sum_{j=1}^{N}
\vec{\varepsilon} _{ij}  {}^{*\!}F_{ij\;\mu\nu} \; ,
\eea
with
\be
{}^{*\!}F_{ij\,\mu \nu} 
=   
\partial_{\mu} B_{ij\, \nu} - \partial_{\nu}B_{ij\, \mu}
- 2\pi \sum_{k=1}^{N}  m _{ij\,k} \Sigma_{k \, \mu \nu}^{(e)} \; ,
\label{eqn:Weyl-DGL-dfs}
\ee
where we have used
\be
\vec{B}_{\mu} = \frac{1}{Ng}\sum_{i=1}^{N}\sum_{j=1}^{N}
\vec{\varepsilon}_{ij}B_{ij \; \mu} \; .
\ee
Now, the dual gauge field consists of $N(N-1)/2$ 
components which parallels the monopole field. 
Reflecting the fact that 
the $N(N-1)/2$ root vectors are not all independent,
the dual gauge field has to satisfy several constraints
connecting different color labels.
Since this redefinition allows us to write
the square of the dual field strength tensor as
\be
({}^{*\!}\vec{F}_{\mu\nu} )^{2}=
\frac{1}{Ng^{2}} \sum_{i=1}^{N} \sum_{j=1}^{N}
({}^{*\!}F_{ij\,\mu \nu} )^{2}
=
\frac{2}{Ng^{2}} \sum_{i<j}^{N}
({}^{*\!}F_{ij\,\mu \nu} )^{2} \; ,
\ee
we find that the DGL theory can be written in  
a manifestly Weyl symmetric form as
\bea
&&
{\cal L}_{\rm DGL} =
\sum_{i < j}^{N}
\left[ - \frac{1}{2 Ng^{2}} {}^{*\!}F_{ij\,\mu \nu}^{2}
+ \left | ( \partial_{\mu} + i B_{ij\,\mu}) \chi_{ij} \right |^{2}
- \lambda \left( |\chi_{ij}|^{2} - v^{2} \right)^{2}   \right].
\label{eqn:Weyl-DGL}
\eea
Clearly, this form is symmetric under the exchange of the color 
labels $i$ or $j$.
This is the extension of our previous work ~\cite{Koma:2000wn} 
from SU(3) to SU($N$).
Apart from the existing constraints, the action takes the form of
a sum of $N(N-1)/2$ 
types of U(1) dual Abelian Higgs (DAH) models.
The Lagrangian density is invariant
under the [U(1)]$^{N(N-1)/2}$ transformation:
\bea
&&
\chi_{ij} \mapsto \chi_{ij}e ^{if_{ij}}, \quad
\chi_{ij}^{*} \mapsto \chi_{ij}^{*} e ^{-if_{ij}}, \quad
B_{ij\;\mu} \mapsto B_{ij\;\mu} - \partial_{\mu} f_{ij},
\nonumber\\*
&&
(i=1,2,\ldots,N \; \mbox{and}  \;
j=i+1,\ldots,N).
\eea
However, this does not imply an increase of symmetry.
The remaining constraints among the dual gauge fields 
express the fact that the dual gauge symmetry 
remains [U(1)]$^{N-1}$.
The number of constraints is found to be 
$N(N-1)/2 - (N-1) =(N-2)(N-1)/2$.
For example, in the SU($N=3$) case we have  
one constraint $B_{12\;\mu}+B_{23\;\mu}
+B_{31\;\mu}=0$~\cite{Koma:2000wn}.

\section{The Generalized Flux-Tube Solutions}

We are interested in the flux-tube solution to be obtained in this 
framework which is associated with external color-electric charges
belonging to some particular representation of SU($N$).
The solution is provided by solving 
the classical field equations derived from the Lagrangian
density for each pair $i < j$:
\bea
&&
\frac{1}{g_{m}^2} \partial^{\mu} {}^{* \!} F_{ij\; \mu \nu}
=
-i \left ( 
\chi_{ij}^{*}\partial_{\nu} \chi_{ij} - \chi_{ij}\partial_{\nu} \chi_{ij}^{*}
\right )
+ 2 B_{ij\;\nu} \chi_{ij}^{*}\chi_{ij}=k_{ij\; \nu} \; ,
\label{eqn:feq-1-m} \\
&&
\left ( \partial_{\mu} +i B_{ij\;\mu} \right )^2 \chi_{ij}
=
 2 \lambda \chi_{ij} ( \chi_{ij}^{*} \chi_{ij} - v^2) \; .
\label{eqn:feq-2-m}
\eea
Here we have defined an ``effective'' dual gauge coupling
\bea
g_{m} \equiv \sqrt{\frac{N}{2}} g .
\eea
Written in terms of the coupling $g_{m}$, 
each pair of the replicated field equations 
has exactly the same form as in the U(1) DGL theory for SU($N=2$).
This is one of the advantages of this formulation.
The flux-tube solution is induced by the 
Dirac string singularity, where the field profiles 
have to behave so as to make the energy of the system finite.
In order to realize this feature more explicitly, it is useful to
parametrize the dual gauge field by two parts as
\bea
B_{ij\;\mu} \equiv B_{ij\;\mu}^{\rm reg}  
+ \sum_{k=1}^{N} m_{ij \; k} B_{k\; \mu}^{\rm sing},
\label{eqn:define-bsing1}
\eea
where the second term,
the singular part of the dual gauge field,
is determined by the relation
\bea
\partial_{\mu}B_{k\; \nu}^{\rm sing}
-\partial_{\nu}B_{k\; \mu}^{\rm sing}
-2 \pi \Sigma_{k\; \mu \nu}^{(e)}= 2\pi C_{k \; \mu \nu}^{(e)}
\label{eqn:define-bsing2}
\eea
with the Coulombic field tensor 
\bea
C_{k\; \mu\nu}^{(e)}(x) 
= \frac{1}{4\pi^{2}} \int d^{4}y \frac{1}{|x-y|^{2}}
{}^{*\!}(\partial \wedge j_{k}^{(e)}(y))_{\mu\nu} \; .
\eea
Note that the square of this field tensor
leads to the interaction Lagrangian between the color-electric
currents via the Coulomb propagator.
It means that this term gives the Coulomb potential for
the static quark-antiquark 
system~\footnote{However, this does not mean that the total
non-confining potential of the flux-tube becomes the Coulomb 
potential, since there are other 
boundary contributions from remaining terms.
In the London limit, the sum of all boundary contributions 
finally leads to the Yukawa 
potential~\cite{Koma:2001pz,Koma:2002rw}.}.
Inserting Eqs.~(\ref{eqn:define-bsing1}) and (\ref{eqn:define-bsing2})
into Eq.~(\ref{eqn:Weyl-DGL-dfs}), the dual field strength tensor
is written as
\bea
{}^{*\!}F_{ij \; \mu\nu} 
= 
\partial_{\mu}B_{ij\; \nu}^{\rm reg}
-\partial_{\nu}B_{ij\; \mu}^{\rm reg}
+2 \pi \sum_{k=1}^{N} m_{ij \; k} C_{k \; \mu\nu}^{(e)}.
\eea
Here both the regular part of the dual gauge field 
$B_{ij\; \mu}^{\rm reg}$ 
and the Coulomb term $C_{k \; \mu\nu}^{(e)}$ 
does not contain any string singularity,
so that the dual field strength tensor does not either.
All the information concerning the string, for instance its position or
length, is now described by the singular part $B_{k\; \mu}^{\rm sing}$,
which is not a part of the gauge field any more.
Since $B_{k\; \mu}^{\rm sing}$ 
directly couples to the monopole field $\chi_{ij}$, 
one can find the general boundary conditions;
on the string the modulus of the monopole field should vanish 
$|\chi_{ij}| =0$, while it takes the value of condensate 
$|\chi_{ij}| \to v$ at a distance from the string.
Simultaneously, the dual gauge field should behave as
$B_{ij\; \nu}^{\rm reg} \to -
\sum_{k=1}^{N} m_{ij \; k} B_{k\; \mu}^{\rm sing}$.

\par
Let us consider the infinitely long flux-tube system with cylindrical and 
translational symmetry along $z$ axis,
in other words, we put a quark at 
$x=y=0$ and $z=-\infty$  and an antiquark at $x=y=0$ and $z=\infty$.
In this case we can neglect the Coulomb term 
$C_{j\;\mu \nu}^{(e)}$.
All fields are now described as 
a function only of the radial coordinate $r$ :
\bea
\chi_{ij} = \phi_{ij}(r) \exp (i \eta_{ij}),
\eea
and
\bea
\bvec{B}_{ij}^{\rm reg}(r) = B_{ij}^{\rm reg}(r) \bvec{e}_{\varphi} 
\equiv  
\frac{\tilde B_{ij}^{\rm reg}(r)}{r} \bvec{e}_{\varphi},
\eea
where $\varphi$ denotes the azimuthal angle around the $z$ axis.
It is useful to note that, 
when the open flux-tube system is considered, 
the phase of the monopole field
$\eta_{ij}$ is single-valued, which 
can be absorbed 
in the regular part of the dual gauge field 
by the redefinition as 
$B_{ij\;\mu}^{\rm reg} + \partial_{\mu} \eta_{ij} \to
B_{ij\;\mu}^{\rm reg}$.
The solution of Eq.~(\ref{eqn:define-bsing2}) 
is found to be
\bea
\sum_{k=1}^{N} m_{ij \; k} \bvec{B}_{k}^{\rm sing}(r)
= -\frac{n_{ij}^{(m)}}{r} \bvec{e}_{\varphi} \; ,
\eea
where
\be
n_{ij}^{(m)}=\sum_{k=1}^{N}m_{ij\; k}n_{k}^{(e)}
= n_{i}^{(e)}-n_{j}^{(e)}
\ee
is the winding number of the flux tube.
The set of integer values 
$\{ n_{1}^{(e)},n_{2}^{(e)},
\ldots,n_{N}^{(e)} \}$
depends on the {\it representation} 
to which the color-electric 
charges placed at the ends of the flux tube belong
\footnote{Thanks to the Weyl symmetric 
formulation, it is not necessary to pay attention to the order 
of winding numbers. 
For example, the bracket
$\{A,B \}$ represents not only $(A,B)$ but also $(B,A)$ (all possible
permutation can be taken).}.
Paying attention to the highest weight 
of a $D$-dimensional representation of the SU($N$) algebra,
being specified by Dynkin indices as 
$[p_{1},p_{2},\ldots,p_{N-1}]$,  we find the simple relation
\be
\{ n_{1}^{(e)},n_{2}^{(e)},\ldots,n_{N}^{(e)} \}=
\{p_{1}, -p_{2}, -(p_{2}+p_{3}),\ldots, -\sum_{i=2}^{N-1}p_{i},0 \}.
\label{eqn:winding-Dynkin}
\ee
Note that since  only $N-1$ of the weight vectors  are independent 
due to $\sum_{j=1}^{N} \vec{w}_{j}=0$,
we can choose $n_{N}^{(e)}=0$ without loss of generality.

\par
The field equations 
(\ref{eqn:feq-1-m}) and (\ref{eqn:feq-2-m}) are now reduced to 
\bea
& &   \frac{d^2 \tilde{B}^{\rm reg}_{ij}}{dr^2} 
- \frac{1}{r}\frac{d \tilde{B}^{\rm reg}_{ij}}{dr}
-2 g_{\rm m}^2 \left ( \tilde{B}^{\rm reg}_{ij} - 
n_{ij}^{(m)} \right ) 
\phi_{ij}^2 = 0 \; , \\
\label{eqn:feq-1-m-cyl}
&&
\frac{d^2\phi_{ij}}{dr^2} +\frac{1}{r}\frac{d\phi_{ij}}{dr}
-
\left ( 
\frac{ \tilde{B}^{\rm reg}_{ij} - n_{ij}^{(m)}}
{r} \right )^2 \phi_{ij}
- 2 \lambda \phi_{ij} (\phi_{ij}^2 - v^2) = 0 \; .
\label{eqn:feq-2-m-cyl}
\eea
Here we can write the boundary conditions for the fields explicitly as
\bea
&&
\tilde{B}^{\rm reg}_{ij} = 0, \quad \phi_{ij} = 
\left  \{
\begin{array}{cc}
0 & (n_{ij}^{(m)}\ne 0)\\
v & (n_{ij}^{(m)}=0)
\end{array}
\right.
\quad {\rm as}\quad r \to 0 \; , \nonumber \\
&&
\tilde{B}^{\rm reg}_{ij} = n_{ij}^{(m)},\quad \phi_{ij} = v 
\qquad\qquad\;\;\; {\rm as}\quad  r \to \infty \; .
\label{eqn:boundary-condition}
\eea
In order to obtain a flux-tube solution, some of
the winding numbers must take non-zero integer values, 
$n_{ij}^{(m)} \ne 0$.
In other words, if $n_{ij}^{(m)}= 0$ for all combinations of $i$ and 
$j$, the field equations have only a trivial solution: 
$\tilde{B}^{\rm reg}_{ij} = 0$, $\phi_{ij} = v$ everywhere.

\section{The generalized string tension}

\par
The energy of the system is, in general, given by 
the spatial integration of the energy-momentum tensor.
Starting from this,
we can specify the {\it string tension} as the energy of the flux tube
per unit length. 
In the system with cylindrical symmetry we have
\bea
\sigma_D 
= 2\pi  \sum_{i < j}^N \int_0^{\infty}rdr
\Biggl [
\frac{1}{2 g_{\rm m}^2} \left ( \frac{1}{r}
\frac{d \tilde B^{\rm reg}_{ij}}{dr} \right )^2
+
\left ( \frac{d \phi_{ij}}{d r} \right)^2
+
\left ( \frac{ \tilde B^{\rm reg}_{i j} 
-  n_{i j}^{(m)}}{r} \right )^2 \! \phi_{ij}^2
+ \lambda ( \phi_{ij}^2 - v^2 )^2 \Biggr ] \; .
\eea
Remarkably, in this system,  the boundary conditions 
of the field profile (\ref{eqn:boundary-condition})
allow to write the string tension as
\bea
\sigma_D
&=&
2\pi v^2 \sum_{i<j}^{N}\left | n_{ij}^{(m)} \right |
+
2\pi \sum_{i<j}^{N} \int_0^{\infty} \!\! r dr \Biggl [
\frac{1}{2g_{\rm m}^2} 
\left ( \frac{1}{r}\frac{d \tilde B^{\rm reg}_{ij}}{dr} 
\pm g_{\rm m}^2 ( \phi_{ij}^2 - v^2 )\right )^2 
\nonumber\\*
&&
+
\left ( \frac{d \phi_{ij}}{d r} 
\pm \left ( \tilde B^{\rm reg}_{ij} -  n_{ij}^{(m)} \right )
\frac{\phi_{ij}}{r} \right)^2 
+
\frac{1}{2}\left (2 \lambda - g_{\rm m}^2 \right ) 
( \phi_{ij}^2 - v^2 )^2 \Biggr ] \; .
\label{eqn:string-tension-2}
\eea
From this expression 
one finds that the so-called Bogomol'nyi 
limit,
\bea
g_{\rm m}^2 = 2\lambda, \quad {\rm or}\quad N g^2 = 4 \lambda,
\label{eqn:bogomolnyi-limit}
\eea
leads to a considerable simplification 
in the evaluation on the string tension.
Note that these relations correspond to taking
$m_{B}=m_{\chi}$ ($\kappa =m_{\chi}/m_{B}=1$).
In this case, the field equations 
turn into first order differential equations:
\bea
&&
\frac{1}{r} \frac{d \tilde B^{\rm reg}_{ij} }{dr}
\pm
g_{\rm m}^2 (\phi_{ij}^2 - v ^2) =0  \; ,
\label{eqn:bogomol-eq-1} \\*
&&
\frac{d\phi_{ij}}{dr} 
\pm 
\left ( \tilde B^{\rm reg}_{ij}  - n_{ij}^{(m)} \right )
\frac{\phi_{ij}}{r}=0 \; .
\label{eqn:bogomol-eq-2}
\eea
If they are satisfied,
the string tension acquires the simple form
\be
\sigma_D
=
2\pi v^2 \sum_{i<j}^{N}\left | n_{ij}^{(m)} \right |
=
2\pi v^2 \sum_{i<j}^{N}\left | n_{i}^{(e)} - n_{j}^{(e)} \right | 
=
2 \pi v^{2} \sum^{N-1}_{k=1} k (N-k)p_{k} \; ,
\label{eqn:st-exact}
\ee
where the final equality holds in view of the relation
between the winding numbers and the Dynkin index 
(\ref{eqn:winding-Dynkin}).
This is the main result of this paper.
The string tension in the fundamental
representation $\sigma_{F}$ is given by
$\sigma_{F} = 2\pi v^{2} (N-1)$ since 
$p_{1}=1$, $p_{i=2,\ldots,N-1}=0$.
For the adjoint representation, $\sigma_{A}=4\pi v^{2}(N-1)$, 
where $p_{1}=p_{N-1}=1$, $p_{i=2,\ldots,N-2}=0$.
It is interesting to note that $\sigma_{A}/\sigma_{F}=2$ for 
arbitrary gauge group SU($N$).
We find that the string tension is proportional to
the number of color-electric Dirac strings inside the flux-tube,
$n_{ij}^{(m)}$. It means that 
the {\it flux counting rule} for the string tension is realized.
This is a good starting point for considerations 
in the type-I ($\kappa <1$)  or type-II ($\kappa >1$) 
dual superconducting vacuum, where
the string tension deviates from the flux counting rule, 
due to the manifest interaction
between the dual gauge field and the monopole field,
in the sense that the corrections will be under control.

\begin{figure}[t]
\includegraphics[ width=17cm]{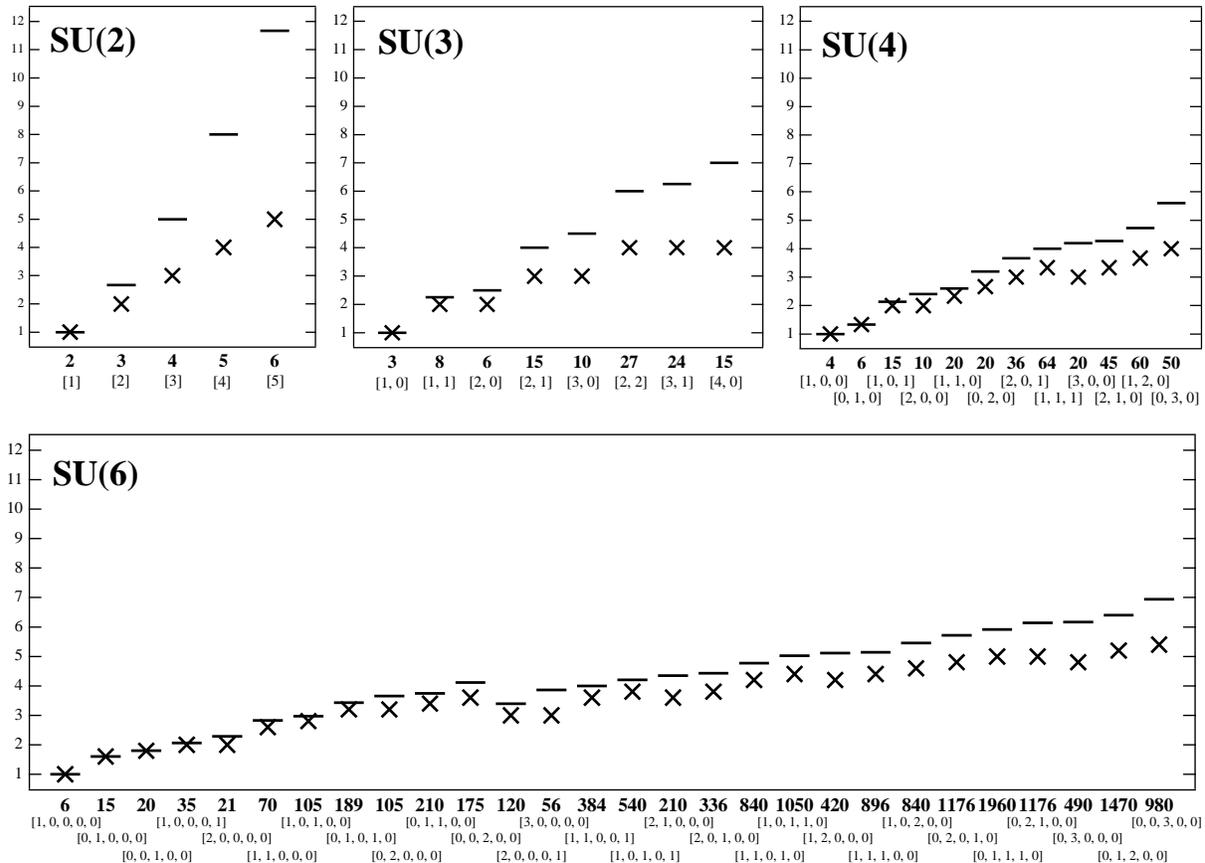}
\caption{The systematic behavior of the 
ratios of  string tensions, 
$d_{D}\equiv \sigma_{D}/\sigma_{F}$,
at the Bogomol'nyi limit $\kappa=1$ (crosses)
associated with color-electric charges in 
various $D$-dimensional representations in 
SU(2), SU(3), SU(4), and SU(6).
The ratio of eigenvalues of quadratic Casimir operators is
shown as black bars.
Boldface numbers and brackets $[p_{1},p_{2},\ldots,p_{N-1}]$
denote the dimension and the Dynkin indices of
each representation, respectively.}
\label{fig:all}
\end{figure}

\par
In Fig.~\ref{fig:all} we show the systematic behavior of 
the ratio between string tensions, 
\bea
d_{D}=\frac{\sigma_{D}}{\sigma_{F}}=
\sum_{k=1}^{N-1}\frac{k(N-k)}{N-1}p_{k} \; ,
\label{eqn:st-exact-ratio}
\eea
for the gauge groups SU(2), SU(3), SU(4), and SU(6).
In comparison, we also plot 
the corresponding ratio between 
the eigenvalues of the quadratic Casimir operator
associated with the 
representations $D$ and $F$,
$C^{(2)}(D)/C^{(2)}(F)$ (see, Appendix~\ref{sec:casimir}).
At glance, one finds that the ratio $d_{D}$ 
tends to approach the Casimir ratio in the large $N$ limit.
This will become clearer in the Appendix~\ref{sec:casimir}, 
where we observe that the  
leading contribution to the ratio 
of eigenvalues of Casimir operator for lower dimensional 
representations is given by
\be
C^{(2)}(D)/C^{(2)}(F) \sim \sum_{k=1}^{N-1}\frac{k(N-k)}{N-1}p_{k} \; ,
\ee
which coincides with Eq.~(\ref{eqn:st-exact-ratio}).

\par
It is amusing to note that if one evaluates
the eigenvalues of quadratic Casimir operator
only from off-diagonal generators, neglecting 
the original
diagonal ones $T_{3}$, $T_{8}$, $\ldots$ , $T_{N^{2}-1}$,
one gets 
\be
C_{\rm offdiag}^{(2)}(D) = \frac{1}{2}\sum_{k=1}^{N-1}k(N-k)p_{k} \; .
\ee
Here only {\it induced} diagonal generators, appearing 
through the commutation relations among off-diagonal generators
are taken into account.
Then, the ratio $C_{\rm offdiag}^{(2)}(D)/C_{\rm offdiag}^{(2)}(F)$
exactly reproduces Eq.~(\ref{eqn:st-exact-ratio}).
Since the DGL theory is based on the Abelian projection scheme, 
where the non-diagonal part of the non-Abelian gauge field is absent,
one might be misled to expect that the string tension of the flux tube
in this approach
scales exclusively with the square of the original diagonal generators.
The result shows, however, that this consideration is too naive. 
The flux tube energy
is mainly controlled by induced diagonal generators.
This clear result
manifestly emerges in the Bogomol'nyi limit.

\section{Summary}

\par
We have formulated the [U(1)]$^{N-1}$ dual Ginzburg-Landau  (DGL)
theory as a low-energy effective theory of 
Abelian-projected SU($N$) gauge theory 
within a manifestly Weyl symmetric procedure.
This leads to a Lagrangian density which corresponds  to
the sum of $N(N-1)/2$ types of U(1) dual Abelian Higgs models 
with $(N-2)(N-1)/2$ constraints among the dual gauge fields.
We have  calculated the string tensions of the flux-tube
solution associated with static charges in various $D$-dimensional 
representations for SU($N$).
This is analytically possible 
at the transition point from a type-I to type-II superconducting 
vacuum, also known as the Bogomol'nyi limit ($\kappa =1$).
We have shown that the string tensions 
satisfy the flux counting rule.
By comparing the ratio of string tensions, $d_{D}\equiv 
\sigma_{D}/\sigma_{F}$, with the ratio of eigenvalues of the 
quadratic Casimir operators, we have found that 
the flux-tube  in the DGL theory mainly
carries the information of induced diagonal generators
rather than that of the original diagonal ones.
This feature leads to the tendency, that 
$d_{D} \to C^{(2)}(D)/C^{(2)}(F)$ in the large $N$ limit.

\par
In this paper we have concentrated on the Bogomol'nyi limit.
The actual type of the dual superconducting vacuum is
not determined yet,  which is one of the 
longstanding problems in this scenario.
Of course, finally it should be determined directly from 
the non-Abelian gauge theory.
The investigation of 
monopole dynamics based on lattice simulations
is a promising way for this purpose.
In fact,  such efforts for the SU(2) and the SU(3)  cases 
are in progress~\cite{Suzuki:2001tp}.
Once the vacuum parameters are found, it becomes  
interesting to compare the ratio of the string tension
of the flux tubes for arbitrary sources in the DGL theory, 
with observed lattice 
data~\cite{Deldar:1999vi,Bali:2000un,Lucini:2001nv,%
DelDebbio:2001kz,DelDebbio:2001sj}
in the non-Abelian theory.

\begin{acknowledgments}
The author is grateful to M. Koma, T. Suzuki and
E.-M. Ilgenfritz for fruitful discussions.
This work is partially supported  by 
the Ministry of Education, Science, Sports and Culture,
Japan, Grant-in-Aid for Encouragement of Young 
Scientists (B), 14740161, 2002.
\end{acknowledgments}

\appendix
\section{Eigenvalue of quadratic Casimir operator}\label{sec:casimir}

The eigenvalue of quadratic Casimir operator $C^{(2)}[D]$ for
arbitrary dimensional representations in SU($N$)
can be expressed in terms of the 
Dynkin indices $[p_{1},p_{2},\ldots,p_{N-1}]$.
For SU(2), SU(3), SU(4), and SU(6) cases:
\bea
\mbox{SU(2):} \quad C^{(2)}[D] &=& 
\frac{p_1}{2}+\left ( \frac{p_{1}}{2}\right )^{2},
\\
\mbox{SU(3):} \quad C^{(2)}[D] &=& 
p_{1}+p_{2} +   \left ( \frac{p_{1}+p_{2}}{2}\right )^{2} 
+\left ( \frac{p_{1}-p_{2}}{2\sqrt{3}}\right )^{2},
\\
\mbox{SU(4):} \quad C^{(2)}[D] &=&\frac{3}{2}(p_{1}+\frac{4}{3}p_{2}+p_{3})
+ \left ( \frac{p_{1}+p_{2}+p_{3}}{2}\right )^{2} 
+\left ( \frac{p_{1}+p_{2}-p_{3}}{2\sqrt{3}}\right )^{2} 
+\left ( \frac{p_{1}-2p_{2}-p_{3}}{2\sqrt{6}}\right )^{2},
\\
\mbox{SU(6):} \quad C^{(2)}[D] &=&
\frac{5}{2}(p_{1}+\frac{8}{5}p_{2}+\frac{9}{5}p_{3}+\frac{8}{5}p_{4}+p_{5})
+\left (\frac{p_{1}+p_{2}+p_{3}+p_{4}+p_{5}}{2}\right )^{2}
\nonumber\\*
&&
+\left (\frac{p_{1}+p_{2}+p_{3}+p_{4}-p_{5}}{2\sqrt{3}}\right )^{2}
+\left (\frac{p_{1}+p_{2}+p_{3}-2p_{4}-p_{5}}{2\sqrt{6}}\right )^{2}
\nonumber\\*
&&
+\left (\frac{p_{1}+p_{2}-3p_{3}-2p_{4}-p_{5}}{2\sqrt{10}}\right )^{2}
+\left (\frac{p_{1}-4p_{2}-3p_{3}-2p_{4}-p_{5}}{2\sqrt{15}}\right )^{2}.
\eea
These formulae can be obtained by paying attention to
the highest weight state of the representation.
As an example, it is instructive to learn the derivation from the SU(2) case.
Let the state which belongs to $D$-dimensional representation
be $|D \rangle$ and its highest weight 
state $|D_{\rm max} \rangle$,
where the latter state is defined so as to satisfy
$E_{+} |D_{\rm max} \rangle=0$ with raising operator
$E_{+} \equiv T_{1} + iT_{2}$.
Operating $T_{3}$ on the highest weight, we get an eigenvalue 
$p_{1}/2$ with the Dynkin index $[p_{1}]$:
\be
T_{3} |D_{\rm max}\rangle = \frac{p_{1}}{2} |D_{\rm max}\rangle.
\ee
Then the eigenvalue of the quadratic Casimir operator is calculated as
\be
C^{(2)}[D] 
=\langle D| \sum_{a=1}^{3} T_{a} T_{a} |D \rangle
=\langle D_{\rm max} | E_{-}E_{+}+T_{3} + T_{3}^{2} |D_{\rm max} \rangle
= \frac{p_{1}}{2}+ \left ( \frac{p_{1}}{2} \right )^{2}.
\ee
This derivation is extended straightforwardly to the SU(3), SU(4), and 
SU(6) cases, which provides the above expressions. 
The dimension $D_{N}$ for the representation 
$[p_{1},p_{2},\ldots, p_{N-1}]$ is 
\bea
D_{N}
&=& \frac{1}{2!  \cdots (N-1)!}(p_{1}+1)(p_{1}+p_{2}+2) \cdots 
(p_{1} + \cdots + p_{N-1} +N-1)\nonumber\\*
&&
\times(p_{2} +1) (p_{2}+p_{3}+ 2)  \cdots 
(p_{2}+ \cdots + p_{N-1} + N-2) 
\nonumber\\*
&&\times 
\cdots
\nonumber\\*
&&
\times(p_{N-2} + p_{N-1} +2)
\nonumber\\*
&&\times 
(p_{N-1}+1).
\eea

\newpage
Explicitly, this means for our cases
\bea
\mbox{SU(2):} \quad D_{2} &=& p_{1}+1,
\\
\mbox{SU(3):} \quad D_{3} &=& 
\frac{1}{2}(p_{1}+1)(p_{1}+p_{2}+2)
\nonumber\\*
&&\times
(p_{2}+1),
\\
\mbox{SU(4):} \quad D_{4}
&=& \frac{1}{2!3!}(p_{1}+1)(p_{1}+p_{2}+2)(p_{1}+p_{2}+p_{3}+3)
 \nonumber\\*
 &&\times 
(p_{2}+1) (p_{2}+p_{3}+2)
\nonumber\\*
&&\times
(p_{3}+1),
\\
\mbox{SU(6):} \quad D_{6} &=&
\frac{1}{2!3!4!5!}
(p_{1}+1)(p_{1}+p_{2}+2)(p_{1}+p_{2}+p_{3}+3)
(p_{1}+p_{2}+p_{3}+p_{4}+4)
(p_{1}+p_{2}+p_{3}+p_{4}+p_{5}+5)
\nonumber\\*
&&\times
(p_{2}+1)(p_{2}+p_{3}+2)(p_{2}+p_{3}+p_{4}+3)
(p_{2}+p_{3}+p_{4}+p_{5}+4)
\nonumber\\*
&&\times
(p_{3}+1)(p_{3}+p_{4}+2)(p_{3}+p_{4}+p_{5}+3)
\nonumber\\*
&&\times
(p_{4}+1)(p_{4}+p_{5}+2) 
\nonumber\\*
&&\times
(p_{5}+1).
\eea


\begin{thebibliography}{10}
\expandafter\ifx\csname bibnamefont\endcsname\relax
  \def\bibnamefont#1{#1}\fi
\expandafter\ifx\csname bibfnamefont\endcsname\relax
  \def\bibfnamefont#1{#1}\fi
\expandafter\ifx\csname url\endcsname\relax
  \def\url#1{\texttt{#1}}\fi
\expandafter\ifx\csname urlprefix\endcsname\relax\def\urlprefix{URL }\fi
\expandafter\ifx\csname bibinfo\endcsname\relax \def\bibinfo#1#2{#2}\fi
\expandafter\ifx\csname eprint\endcsname\relax \def\eprint#1{#1}\fi

\bibitem{tHooft:1975pu}
\bibinfo{author}{\bibfnamefont{G.}~\bibnamefont{'t~Hooft}}, in
  \emph{\bibinfo{booktitle}{High-Energy Physics. Proceedings of the EPS
  International Conference, Palermo, Italy, 23-28 June 1975, Vol. 2}}, edited
  by \bibinfo{editor}{\bibfnamefont{A.}~\bibnamefont{Zichichi}}
  (\bibinfo{publisher}{Bologna}, \bibinfo{year}{1976}), pp.
  \bibinfo{pages}{1225--1249}.

\bibitem{Mandelstam:1974pi}
\bibinfo{author}{\bibfnamefont{S.}~\bibnamefont{Mandelstam}},
  \bibinfo{journal}{Phys. Rept.} \textbf{\bibinfo{volume}{23C}},
  \bibinfo{pages}{245} (\bibinfo{year}{1976}).

\bibitem{Suzuki:1988yq}
\bibinfo{author}{\bibfnamefont{T.}~\bibnamefont{Suzuki}},
  \bibinfo{journal}{Prog. Theor. Phys.} \textbf{\bibinfo{volume}{80}},
  \bibinfo{pages}{929} (\bibinfo{year}{1988}).

\bibitem{Maedan:1988yi}
\bibinfo{author}{\bibfnamefont{S.}~\bibnamefont{Maedan}} \bibnamefont{and}
  \bibinfo{author}{\bibfnamefont{T.}~\bibnamefont{Suzuki}},
  \bibinfo{journal}{Prog. Theor. Phys.} \textbf{\bibinfo{volume}{81}},
  \bibinfo{pages}{229} (\bibinfo{year}{1989}).

\bibitem{Suganuma:1995ps}
\bibinfo{author}{\bibfnamefont{H.}~\bibnamefont{Suganuma}},
  \bibinfo{author}{\bibfnamefont{S.}~\bibnamefont{Sasaki}}, \bibnamefont{and}
  \bibinfo{author}{\bibfnamefont{H.}~\bibnamefont{Toki}},
  \bibinfo{journal}{Nucl. Phys.} \textbf{\bibinfo{volume}{B435}},
  \bibinfo{pages}{207} (\bibinfo{year}{1995}), \eprint{hep-ph/9312350}.

\bibitem{Sasaki:1995sa}
\bibinfo{author}{\bibfnamefont{S.}~\bibnamefont{Sasaki}},
  \bibinfo{author}{\bibfnamefont{H.}~\bibnamefont{Suganuma}}, \bibnamefont{and}
  \bibinfo{author}{\bibfnamefont{H.}~\bibnamefont{Toki}},
  \bibinfo{journal}{Prog. Theor. Phys.} \textbf{\bibinfo{volume}{94}},
  \bibinfo{pages}{373} (\bibinfo{year}{1995}).

\bibitem{tHooft:1981ht}
\bibinfo{author}{\bibfnamefont{G.}~\bibnamefont{'t~Hooft}},
  \bibinfo{journal}{Nucl. Phys.} \textbf{\bibinfo{volume}{B190}},
  \bibinfo{pages}{455} (\bibinfo{year}{1981}).

\bibitem{Ezawa:1982bf}
\bibinfo{author}{\bibfnamefont{Z.~F.} \bibnamefont{Ezawa}} \bibnamefont{and}
  \bibinfo{author}{\bibfnamefont{A.}~\bibnamefont{Iwazaki}},
  \bibinfo{journal}{Phys. Rev.} \textbf{\bibinfo{volume}{D25}},
  \bibinfo{pages}{2681} (\bibinfo{year}{1982}).

\bibitem{Ezawa:1982ey}
\bibinfo{author}{\bibfnamefont{Z.~F.} \bibnamefont{Ezawa}} \bibnamefont{and}
  \bibinfo{author}{\bibfnamefont{A.}~\bibnamefont{Iwazaki}},
  \bibinfo{journal}{Phys. Rev.} \textbf{\bibinfo{volume}{D26}},
  \bibinfo{pages}{631} (\bibinfo{year}{1982}).

\bibitem{Kronfeld:1987vd}
\bibinfo{author}{\bibfnamefont{A.~S.} \bibnamefont{Kronfeld}},
  \bibinfo{author}{\bibfnamefont{G.}~\bibnamefont{Schierholz}},
  \bibnamefont{and} \bibinfo{author}{\bibfnamefont{U.~J.} \bibnamefont{Wiese}},
  \bibinfo{journal}{Nucl. Phys.} \textbf{\bibinfo{volume}{B293}},
  \bibinfo{pages}{461} (\bibinfo{year}{1987}).

\bibitem{Suzuki:1990gp}
\bibinfo{author}{\bibfnamefont{T.}~\bibnamefont{Suzuki}} \bibnamefont{and}
  \bibinfo{author}{\bibfnamefont{I.}~\bibnamefont{Yotsuyanagi}},
  \bibinfo{journal}{Phys. Rev.} \textbf{\bibinfo{volume}{D42}},
  \bibinfo{pages}{4257} (\bibinfo{year}{1990}).

\bibitem{Brandstater:1991sn}
\bibinfo{author}{\bibfnamefont{F.}~\bibnamefont{Brandstaeter}},
  \bibinfo{author}{\bibfnamefont{U.~J.} \bibnamefont{Wiese}}, \bibnamefont{and}
  \bibinfo{author}{\bibfnamefont{G.}~\bibnamefont{Schierholz}},
  \bibinfo{journal}{Phys. Lett.} \textbf{\bibinfo{volume}{B272}},
  \bibinfo{pages}{319} (\bibinfo{year}{1991}).

\bibitem{Shiba:1995db}
\bibinfo{author}{\bibfnamefont{H.}~\bibnamefont{Shiba}} \bibnamefont{and}
  \bibinfo{author}{\bibfnamefont{T.}~\bibnamefont{Suzuki}},
  \bibinfo{journal}{Phys. Lett.} \textbf{\bibinfo{volume}{B351}},
  \bibinfo{pages}{519} (\bibinfo{year}{1995}), \eprint{hep-lat/9408004}.

\bibitem{Bali:1996dm}
\bibinfo{author}{\bibfnamefont{G.~S.} \bibnamefont{Bali}},
  \bibinfo{author}{\bibfnamefont{V.}~\bibnamefont{Bornyakov}},
  \bibinfo{author}{\bibfnamefont{M.}~\bibnamefont{Muller-Preussker}},
  \bibnamefont{and}
  \bibinfo{author}{\bibfnamefont{K.}~\bibnamefont{Schilling}},
  \bibinfo{journal}{Phys. Rev.} \textbf{\bibinfo{volume}{D54}},
  \bibinfo{pages}{2863} (\bibinfo{year}{1996}), \eprint{hep-lat/9603012}.

\bibitem{Bornyakov:2001nd}
\bibinfo{author}{\bibfnamefont{V.}~\bibnamefont{Bornyakov}} \emph{et~al.},
  \bibinfo{journal}{Nucl. Phys. Proc. Suppl.} \textbf{\bibinfo{volume}{106}},
  \bibinfo{pages}{634} (\bibinfo{year}{2002}), \eprint{hep-lat/0111042}.

\bibitem{Koma:2000wn}
\bibinfo{author}{\bibfnamefont{Y.}~\bibnamefont{Koma}} \bibnamefont{and}
  \bibinfo{author}{\bibfnamefont{H.}~\bibnamefont{Toki}},
  \bibinfo{journal}{Phys. Rev.} \textbf{\bibinfo{volume}{D62}},
  \bibinfo{pages}{054027} (\bibinfo{year}{2000}), \eprint{hep-ph/0004177}.

\bibitem{Koma:2001ut}
\bibinfo{author}{\bibfnamefont{Y.}~\bibnamefont{Koma}},
  \bibinfo{author}{\bibfnamefont{E.~M.} \bibnamefont{Ilgenfritz}},
  \bibinfo{author}{\bibfnamefont{H.}~\bibnamefont{Toki}}, \bibnamefont{and}
  \bibinfo{author}{\bibfnamefont{T.}~\bibnamefont{Suzuki}},
  \bibinfo{journal}{Phys. Rev.} \textbf{\bibinfo{volume}{D64}},
  \bibinfo{pages}{011501} (\bibinfo{year}{2001}), \eprint{hep-ph/0103162}.

\bibitem{Koma:2000hw}
\bibinfo{author}{\bibfnamefont{Y.}~\bibnamefont{Koma}},
  \bibinfo{author}{\bibfnamefont{E.~M.} \bibnamefont{Ilgenfritz}},
  \bibinfo{author}{\bibfnamefont{T.}~\bibnamefont{Suzuki}}, \bibnamefont{and}
  \bibinfo{author}{\bibfnamefont{H.}~\bibnamefont{Toki}},
  \bibinfo{journal}{Phys. Rev.} \textbf{\bibinfo{volume}{D64}},
  \bibinfo{pages}{014015} (\bibinfo{year}{2001}), \eprint{hep-ph/0011165}.

\bibitem{Bogomolny:1976de}
\bibinfo{author}{\bibfnamefont{E.~B.} \bibnamefont{Bogomol'nyi}},
  \bibinfo{journal}{Sov. J. Nucl. Phys.} \textbf{\bibinfo{volume}{24}},
  \bibinfo{pages}{449} (\bibinfo{year}{1976}).

\bibitem{deVega:1976mi}
\bibinfo{author}{\bibfnamefont{H.~J.} \bibnamefont{de~Vega}} \bibnamefont{and}
  \bibinfo{author}{\bibfnamefont{F.~A.} \bibnamefont{Schaposnik}},
  \bibinfo{journal}{Phys. Rev.} \textbf{\bibinfo{volume}{D14}},
  \bibinfo{pages}{1100} (\bibinfo{year}{1976}).

\bibitem{Chernodub:1999xi}
\bibinfo{author}{\bibfnamefont{M.~N.} \bibnamefont{Chernodub}},
  \bibinfo{journal}{Phys. Lett.} \textbf{\bibinfo{volume}{B474}},
  \bibinfo{pages}{73} (\bibinfo{year}{2000}), \eprint{hep-ph/9910290}.

\bibitem{Suzuki:2001tp}
\bibinfo{author}{\bibfnamefont{T.}~\bibnamefont{Suzuki}} \emph{et~al.},
  \bibinfo{journal}{Nucl. Phys. Proc. Suppl.} \textbf{\bibinfo{volume}{106}},
  \bibinfo{pages}{631} (\bibinfo{year}{2002}), \eprint{hep-lat/0110059}.

\bibitem{Deldar:1999vi}
\bibinfo{author}{\bibfnamefont{S.}~\bibnamefont{Deldar}},
  \bibinfo{journal}{Phys. Rev.} \textbf{\bibinfo{volume}{D62}},
  \bibinfo{pages}{034509} (\bibinfo{year}{2000}), \eprint{hep-lat/9911008}.

\bibitem{Bali:2000un}
\bibinfo{author}{\bibfnamefont{G.~S.} \bibnamefont{Bali}},
  \bibinfo{journal}{Phys. Rev.} \textbf{\bibinfo{volume}{D62}},
  \bibinfo{pages}{114503} (\bibinfo{year}{2000}), \eprint{hep-lat/0006022}.

\bibitem{Lucini:2001nv}
\bibinfo{author}{\bibfnamefont{B.}~\bibnamefont{Lucini}} \bibnamefont{and}
  \bibinfo{author}{\bibfnamefont{M.}~\bibnamefont{Teper}},
  \bibinfo{journal}{Phys. Rev.} \textbf{\bibinfo{volume}{D64}},
  \bibinfo{pages}{105019} (\bibinfo{year}{2001}), \eprint{hep-lat/0107007}.

\bibitem{DelDebbio:2001kz}
\bibinfo{author}{\bibfnamefont{L.}~\bibnamefont{Del~Debbio}},
  \bibinfo{author}{\bibfnamefont{H.}~\bibnamefont{Panagopoulos}},
  \bibinfo{author}{\bibfnamefont{P.}~\bibnamefont{Rossi}}, \bibnamefont{and}
  \bibinfo{author}{\bibfnamefont{E.}~\bibnamefont{Vicari}},
  \bibinfo{journal}{Phys. Rev.} \textbf{\bibinfo{volume}{D65}},
  \bibinfo{pages}{021501} (\bibinfo{year}{2002}), \eprint{hep-th/0106185}.

\bibitem{DelDebbio:2001sj}
\bibinfo{author}{\bibfnamefont{L.}~\bibnamefont{Del~Debbio}},
  \bibinfo{author}{\bibfnamefont{H.}~\bibnamefont{Panagopoulos}},
  \bibinfo{author}{\bibfnamefont{P.}~\bibnamefont{Rossi}}, \bibnamefont{and}
  \bibinfo{author}{\bibfnamefont{E.}~\bibnamefont{Vicari}},
  \bibinfo{journal}{JHEP} \textbf{\bibinfo{volume}{01}}, \bibinfo{pages}{009}
  (\bibinfo{year}{2002}), \eprint{hep-th/0111090}.

\bibitem{Koma:2001pz}
\bibinfo{author}{\bibfnamefont{Y.}~\bibnamefont{Koma}},
  \bibinfo{author}{\bibfnamefont{M.}~\bibnamefont{Koma}},
  \bibinfo{author}{\bibfnamefont{D.}~\bibnamefont{Ebert}}, \bibnamefont{and}
  \bibinfo{author}{\bibfnamefont{H.}~\bibnamefont{Toki}},
  \emph{\bibinfo{title}{Towards the string representation of the dual Abelian
  Higgs model beyond the London limit}} (\bibinfo{year}{2001}),
  \eprint{hep-th/0108138}.

\bibitem{Koma:2002rw}
\bibinfo{author}{\bibfnamefont{Y.}~\bibnamefont{Koma}},
  \bibinfo{author}{\bibfnamefont{M.}~\bibnamefont{Koma}},
  \bibinfo{author}{\bibfnamefont{D.}~\bibnamefont{Ebert}}, \bibnamefont{and}
  \bibinfo{author}{\bibfnamefont{H.}~\bibnamefont{Toki}},
  \emph{\bibinfo{title}{Effective string action for the U(1) x U(1) dual
  Ginzburg- Landau theory beyond the London limit}} (\bibinfo{year}{2002}),
  \eprint{hep-th/0206074}.

\end{thebibliography}

\end{document}